\pdfoutput=1
\documentclass[11pt]{article}
\usepackage{geometry}
\usepackage[hidelinks]{hyperref}
\usepackage{graphicx}
\usepackage{comment}
\usepackage{xspace}
\usepackage{color}
\usepackage[hang,flushmargin]{footmisc} 
\usepackage{todonotes}
\usepackage[normalem]{ulem}
\usepackage{epstopdf} 
\usepackage{wrapfig}
\usepackage{subcaption}
\usepackage{csquotes}
\RequirePackage{amsmath,amsfonts,amssymb}
\RequirePackage{graphicx,xcolor}
\RequirePackage{booktabs}
\newcommand{\chaterji}[1]{\textcolor{blue}{SC: #1}}

\newcommand{\parinaz}[1]{\textcolor{blue}{PN: #1}}

\newcommand{\eg}{{\em e.g.},\xspace}
\newcommand{\ie}{{\em i.e.},\xspace}

\begin{document}
\title{Resilient Cyberphysical Systems and their Application Drivers: A Technology Roadmap}
\date{}
\maketitle
{\centering \large \noindent {\bf Authors}: Somali Chaterji, Parinaz Naghizadeh, Muhammad Ashraful Alam, Saurabh Bagchi, Mung Chiang, David Corman, Brian Henz, Suman Jana, Na Li, Shaoshuai Mou, Meeko Oishi, Chunyi Peng, Tiark Rompf, Ashutosh Sabharwal, Shreyas Sundaram, James Weimer, Jennifer Weller.}

\bigskip

\section{Abstract}
\label{sec:abstract}

Cyberphysical systems (CPS) are ubiquitous in our personal and professional lives, and they promise to dramatically improve micro-communities (\eg urban farms, hospitals), macro-communities (\eg cities and metropolises), urban structures (\eg smart homes and cars), and living structures (\eg human bodies, synthetic genomes). 
The question that we address in this article pertains to designing these CPS systems to be resilient-from-the-ground-up, and in ``learning'' iterations, resilient-by-reaction. An optimally designed system is resilient to both unique and recurrent attacks with a minimal overhead from ``fitting''. 
In the following, our focus is on design and deployment innovations that are broadly applicable across a range of application areas and drivers.

We divide our paper into three broad themes. \underline{First}, we present \textbf{three prominent application drivers} that can lay the basis for the existing and emerging technologies, as follows: \textit{smart cities and digital agriculture}; \textit{planet-scale IoT}; and \textit{internet-of-medical-things (IoMT)}.
We select the three application drivers as representative examples of different possible application scenarios that a CPS ``designer'' may face. These  scenarios are orthogonal to each other, operate at different scales, and collectively cover key application domains. Concretely, while the scale of smart cities and digital agriculture on small to large-sized farms can be large, they are not nearly as large as planet-scale IoT, where the scale may pose unique challenges to start with, requiring the consideration of the scale from the outset. For example, a protocol that relies on lots of messages being passed among the participating nodes may work well from an energy efficiency and timeliness standpoint when all the nodes are nearby, say connected through Bluetooth Low Energy (BLE) links. However, when the system is planet-scale and the nodes are dispersed, potentially over large geographic regions, or in highly congested environments, such as urban road environments, such a protocol will become infeasible. 
In contrast, for IoMT, there is a unique set of challenges with some critical safety requirements as they relate to individuals in medical scenarios. 

\underline{Second}, we lay out the foundational technologies---hardware and algorithms---which will power resilient CPS. We base this discussion on two complementary threads for imbuing resilience in these systems, namely, \textit{resilience-by-design} and \textit{resilience-by-reaction}.
Overall, the notion of resilience can be thought of in the light of three main sources of lack of resilience, as follows: \textit{exogenous factors}, such as natural variations and attack scenarios; \textit{mismatch between engineered designs and exogenous factors} ranging from DDoS (distributed denial-of-service) attacks or other cybersecurity nightmares, so called ``black swan'' events, disabling critical services of the municipal electrical grids and other connected infrastructures, data breaches, and network failures; and the \textit{fragility of engineered designs themselves} encompassing bugs, human-computer interactions (HCI), and the overall complexity of real-world systems.  
One factor to bear in mind is that with successive iterations,  algorithms that are initially resilient-by-reaction become resilient-by-design. This is because as algorithms ``learn'' to react to perturbations, the future versions of the algorithms are \textit{designed} to react. An analogy may be made here to 4G and 5G wireless network technologies, where initially 4G was resilient-by-reaction but now 5G is more resilient-by-design. 

\underline{Third}, in our concluding theme, we come up with a snapshot of research-based and production-based emerging trends to set the tone and the possibilities for evolution of resilient CPS. 

\section{Introduction}
\label{sec:introduction}

\noindent

\noindent A cyberphysical system (CPS) consists of a variety of hardware and software components that interact with one another and with the physical environment in which the system is deployed. The hardware and the software can be altered intentionally (by external actors) or unintentionally (physical failures, bugs, mis-configuration, or simply stemming from the system's complexity) leading to consequences unanticipated by the original system designers. A resilient CPS must be able to bounce back to a functional state, \textit{and not simply cease functioning}, when faced with these unanticipated events, and further learn from them and evolve to protect against future failures.

Our goal in this paper is to lay out a vision for resilience in interconnected CPS for short-term and mid-term research challenges to realize the vision. Among a variety of possible options, we will illustrate challenges and opportunities by considering the hardware and software features associated with different scales and forms of application drivers, ranging from IoMT to planet-scale IoT. These topics fall in different parts of the spectrum of tradeoffs between design complexity \textit{vs}. security needs, societal \textit{vs}. individual tolerance, and our willingness to bear the costs of resilience. 
From a more foundational resilience standpoint, we frame our discussion around two complementary approaches, as follows: 

\begin{enumerate}
    \item {\em Resilience-by-design}: This approach designs and develops CPS with resilience in mind. The process involves careful, and ideally verified, software development methodologies, proper configuration of the systems, and hardware verification for the sensors and the actuators. While this approach can typically make the system resilient against what can be predicted from existing knowledge, it remains blind to future events. Further, in this approach, overly engineered solutions will be counter-productive because some (extrapolated) problems may never arise in practice and thus will make the system overfit; such overfitting will likely make the system more fragile in the face of unanticipated events. We elaborate on how resilience-by-design should account for these challenges. 
    
    \item {\em Resilience-by-reaction}: This is the approach that deals with run-time {\em perturbations} and imbues the systems with the ability to recover quickly after the occurrence of any such perturbation. It involves detecting errors and reconfiguring the system in the face of events that occur at run-time rather than (potentially) overfitting during the system's design and engineering. 
\end{enumerate}

We also make specific the notion of {\em \bf perturbations} that we want to deal with. These take three forms: \textit{natural} software or hardware failures (due to bugs, aging, misconfigurations, etc.), \textit{maliciously induced} failures or security attacks (either from outside the system or from an insider such as a system operator), and \textit{unexpected inputs} (\eg autonomous systems will have to deal with the physical environment with unexpected inputs, such as natural variations). In addition, as noted before, fragility of carefully engineered designs (without taking into consideration resilience measures, such as graceful degradation) in the face of HCI or just the overall scale and complexity of systems may result in failures.

We conclude by discussing the application drivers for the evolution of CPS technologies starting with smart cities (and digital agriculture), moving on to planet-scale applications, and finally to the emerging class of ``Internet-of-Medical-Things'' (IoMT), and more broadly, snapshots of bio-IoT; the latter in the emerging technologies section.

\section{Resilience by Design}
\label{sec:design}

Here we lay out the design principles for resilience that exist today to some extent and are being actively researched. These include both software and hardware development, with a focus on what is special to CPS.

    
\subsection{Fog computing for real-time CPS}
Fog networking is the emerging architecture that is envisioned to bring an array of  functionalities, including computation, communication, and control, closer to end-users by leveraging the availability of resources on the Cloud-to-Things continuum \cite{chiang2016fog}. The shift to this architecture, as opposed to traditional centralized and cloud-based architectures, is motivated by the increasing variety of edge devices that are available today, ranging from wearables to smart phones and tablets to drones. In particular, the  abundance of these devices presents opportunities for designing innovative architectures that can enable real-time operation and monitoring of CPS, while at the same time introducing new challenges in ensuring reliability and resilience of such CPS. We detail some of these opportunities and challenges below.

On one hand, the wide availability of edge devices presents us with an opportunity to re-design or update existing CPS architectures to  improve their resilience. In particular, available resources on edge devices can be used as additional redundancies in existing CPS. For instance, drones or other UAVs can be made available on demand to manage unexpected loads on the communication infrastructure, or to temporarily replace a faulty base station. In general, these devices could supplement both software/control as well as hardware/physical resources of an existing physical or cyberphysical system. Fog resources can be further leveraged in real-time to monitor existing (cyber)physical systems and ensure reliable operation. Exploring such opportunities for novel resilient designs will require research into real-time integration of fog resources with the software and hardware of existing CPS architectures, as well as \textit{a priori} planning to ensure the availability of sufficient fog resources on demand. 

On the other hand, while the ubiquity of fog resources enables novel CPS architectures, the edge devices constituting these designs vary vastly in their capabilities and reliability. Most notably, the energy constraints and mobility of these resources, as well as the increased difficulty in securing these devices against attackers, introduce new challenges in resilient design. These challenges should be taken into account to ensure uninterrupted operation and control of CPS that are reliant on a fog networking architecture. 
\subsection{Multi-agent systems for secure CPS} 
A multi-agent network is a collection of connected, interactive agents, each of which is capable of local sensing, computing, communicating and/or actuating \cite{MS15DC}. By working as a cohesive whole, multi-agent networks are usually able to accomplish missions that are well beyond the capabilities of individual agents, and offer expanded capabilities for recognized military uses as well as a wide variety of civilian uses; \eg search and rescue, cargo delivery, scientific data collection, and homeland security operations \cite{FJS09DC}. In contrast to networks with a centralized coordinator, multi-agent networks operate in a \emph{distributed} way, which achieve global objectives through only local coordination among nearby agents \cite{RJR07PIEEE}. Advantages of multi-agent networks equipped with such distributed algorithms include: \emph{scalability} in the sense that coordination among agents scales well with the network size, \emph{robustness} in the sense that individual agent/link failures usually do not affect functionality of the multi-agent network as long as the network is connected, and \emph{efficiency} in the sense that a globally complex objective could be decomposed into locally simpler sub-objectives that could be executed in parallel by all agents in the network \cite{MM10DC}.

The strong dependence of distributed algorithms on local coordination among nearby agents also raises a major concern that sophisticated cyber-attacks such as Byzantine attacks~\cite{LeBlanc12PHD}, may crash the entire network through compromising one or more vulnerable agents~\cite{FFF13TAC}. Such attacks could target the confidentiality, integrity and availability (CIA) of transmitted data, which are targets usually favored by attackers, especially in the large and open environment where practical multi-agent networks such as electrical grids operate~\cite{SAM11PIEEE}. It is then of research significance to understand how to achieve \emph{resilience} in order for multi-agent networks to continue satisfactory performance in hostile environment. Major challenges include: 
\begin{itemize}

\item \emph{Fully distributed scenario.} Each agent in large-scale multi-agent networks is only provided with locally accessible information, which makes it challenging to identify and then isolate sophisticated attacks.  
\item \emph{Sophisticated attacks.} Cyber-attacks are now becoming increasingly sophisticated, for example exploiting insider credentials to obtain data access without being detected for a long time, using approaches to obtaining that information using techniques such as spam/phishing/malware. Traditional cyber security perimeters have dissolved for which static cyber defense tools are not sufficient any more.  
\item \emph{Real-time response.} Large-scale multi-agent networks also introduce a large attack surface, for which massive attacks could be launched simultaneously at multiple agents. Impacts of these massive attacks can also propagate rapidly across the whole multi-agent networks because of the coordinated autonomy among agents.  
\item \emph{Highly dynamic environment.} Dynamics of operational environment requires resilient strategies for multi-agent networks to function in time-varying environments, especially for time-varying networks, time-varying attackers with high mobility, and so on.
\end{itemize}

Approaches to address the above challenges for resilience of future multi-agent systems will be an integration of advanced techniques such as AI and ML (which are capable of processing large volumes of data online) with classical theories in control and optimization (which are designed for optimal and autonomous actions). Promising research directions include: 
\begin{itemize}
\item Development of secured distributed optimization algorithms~\cite{Shreyas13Selected,SB19TAC,XSS19NACO}. One way to achieve this is by integration of game theory and online learning into existing distributed optimization rules, under which each agent is able to estimate the attack policy of an associated strategic attacker and autonomously compensate for the attackers' actions. 
\item Introduction of additional layers of neural networks at each agent for identification and isolation of cyber-attacks. Well-trained neural networks have the advantage of dealing with typical attacks with known modes. Unsupervised learning methods could be introduced for identifying agents' anomalous behaviors. This could be further classified to be attacks or not with human corrections, which in turn could enlarge the data samples for training neural networks. 
\item Integration of reinforcement learning at each agent to enable agents' performance improvement over time.
To sum up, future research for resilience of multi-agent networks need to be \emph{effective} in detecting, analyzing, and preventing sophisticated threats, \emph{efficient} in terms of automatically providing faster solutions over traditional approaches, \emph{adaptive} to highly dynamic environments, and \emph{scalable} in terms of only using local information and coordination among nearby neighbor agents.
\end{itemize}

\subsection{Game-theoretic modeling and investment} 
Game theory provides a mathematical framework to reason about outcomes in scenarios containing multiple decision-makers (or players), each of whom has an available set of actions.  Each player attempts to maximize its own utility, which is a function of the actions taken by all players.  
As such, game theory is a natural framework to systematically reason about the design of resilient systems, where the players can represent a combination of different system designers (each potentially managing a different part of the system, and with their own interests), and adversaries who attempt to disrupt the system.  

As game theory becomes increasingly applied in complex CPS, there are many challenges that remain to be tackled.  First, CPS typically involve a large number of interdependent subsystems, along with multiple defenders and attackers.  Understanding the interplay between the interdependencies and the game-theoretic decisions made by the defenders is, in general, a difficult task.  There has been a large amount of work on such {\it interdependent security games} for various classes of attacks and network topologies \cite{laszka2015survey,miura2008security, nguyen2009stochastic,varian2004system,kovenock2018optimal,hota2016interdependent,brown2018security}, but much remains to be done to move towards more realistic attack models.  Along the same lines, the problem of {\it designing} the interdependencies between the subsystems is also an important challenge, as this will have important implications for the resilience of the overall system; recent work along these lines include \cite{goyal2014attack,blume2011network,schwartz2011network,hota2016optimal,heinecke2018resilience}.  Furthermore, in large-scale CPS with  many decision-makers, each individual player will generally have access to information that the other players do not have.  While such scenarios are typically considered under the framework of Bayesian games \cite{osborne2004introduction}, the interplay between the dynamic systems that form the CPS and the control actions that each decision-maker must apply (in the presence of adversaries or disturbances) introduces substantial challenges.  Recent efforts along these lines include \cite{gupta2016dynamic,gupta2014common,zhu2011stackelberg,miao2018hybrid}.

In addition to the above game-theoretic models that consider only the interactions between strategic (rational) players, it is also important to incorporate elements of randomness, faults, and disturbances when designing for resilience \cite{zhu2015game}.   For example, recent work has considered network design problems where edges can be removed by a rational attacker with a certain probability, and fail on their own with another probability \cite{schwartz2011network}.  Thus, the defender must now account for both strategic and random disruptions to their system when planning their investments.  Various approaches have been developed over the past several years to extend classical game-theoretic equilibrium concepts to encompass irrational/malicious players, or coalitions of strategic players \cite{moscibroda2006selfish,blum2008regret,halpern2008beyond}.  There has also been a recent effort to introduce mathematical models of human decision-making (drawn from the behavioral economics and psychology literature) into a game-theoretic framework for security in interdependent systems \cite{hota2018gametheoretic,abdallah2019behavior}.

\subsection{Formal proofs for secure ML-driven CPS} 
Complex CPS are increasingly incorporating  Machine Learning (ML) models as part of their decision-making process. ML allows CPS to seamlessly adapt to a wide variety of environments. However, state-of-the-art ML architectures like neural networks are highly non-linear and hard to interpret. Therefore, these models tend of make erroneous predictions for rare inputs (\eg subtle human-imperceptible changes in image pixels) which are often not seen during testing~\cite{szegedy2013intriguing}. In safety and security-critical settings like self-driving cars, autonomous ships, etc, such errors can have serious implications for the system's resilience, leading to  disastrous consequences including fatality.

One way of minimizing such risks is to check whether a ML model satisfies a set of  desirable safety/security properties. For example, any perturbation in an input image's pixels bounded by some $L_\infty$ norm should not change the final prediction.  Formal verification is a popular technique for checking safety and security properties in traditional software.

However, most existing  traditional  verification  techniques do not apply to popular ML models like deep neural networks due  to their complexity and non-convexity. Successful verification approaches for traditional software like  Satisfiability Modulo Theory (SMT) solvers  do  not  scale  beyond  toy  ML  models  because  they are  not  designed  to  handle  floating  point  values  or  transcendental  functions  (e.g.,  exp,  tanh), ubiquitous in modern ML models like DNNs.  Even the problem of verifying simple safety properties of a neural networks with piecewise-linear activation functions like Relus is NP-complete~\cite{katz2017reluplex}. The key challenge in verifying neural networks is to soundly overestimate the output (i.e., not missing any possible output) of a network for a given range of inputs. Several recent  techniques for soundly overestimating the output of the network that have shown initial promise include 
symbolic intervals~\cite{reluval2018,shiqi2018efficient}, abstract domains~\cite{gehrai,singh2018fast}, convex dual~\cite{wong2018provable,dvijotham2018dual}, and Lipschitz-continuity-based-approaches~\cite{weng2018towards,zhang2018recurjac,zhang2018crown}, SDP~\cite{raghunathan2018semidefinite}. 

However, the scalability of verification techniques to large networks used in real-world CPS still remains a significant challenge. Moreover, all current neural network verification tools focus on verifying safety properties on input regions around a limited set of test data points. The fundamental assumption behind such approaches is that the observation on individual test data points will generalize to unseen data points. We believe this fundamental question requires making certain probabilistic assumptions on the distributions of the datasets (i.e., the distribution of the test data points is the same as that of the unseen ones). A clear understanding of such assumptions is extremely important for understanding the assurance provided by the verification process. Finally, supporting a richer set of safety/security properties (e.g., a self-driving car must drive similarly on the same road under different realistic lighting conditions) is also crucial for real-world deployment of ML verification techniques. 

\subsection
{Model-based reinforcement learning (RL)} 
The idea underlying RL is that the agent will learn from the environment by repeated interactions with the environment and receiving rewards or penalties for performing actions. This is particularly relevant for resilient CPS as the system may be embedded in an uncertain physical environment, and it will have to ``learn'' the set of actions for resilient operation. 
The RL process can be modeled as a loop that works as follows:

	Agent receives state $S0$ from the environment; based on that state $S0$, agent takes an action $A0$; environment transitions to a new state $S1$; environment gives some reward $R1$ to the agent; continues iteratively till some termination condition is reached, \eg time limit is reached or the reward is high enough. 
	
This RL loop outputs a sequence of states, actions, and rewards. The goal of the agent is to maximize the expected cumulative reward. The cumulative reward at a time step \textit{t} looking to the future over a bounded horizon till $\Theta+1$ can be written as:
\begin{align*}
	G_t = \sum_{k=0}^{\Theta} \gamma^k R_{t+k+1} \text{ where } \gamma  \in [0,1)
\end{align*}

The parameter $\gamma$ controls the tradeoff between immediate versus long-term rewards. The larger the $\gamma$, the learning agent cares more about the long-term rewards. Conversely, the smaller the $\gamma$, our agent cares more about the short-term rewards.


We can think of RL-based algorithms answering three kinds of questions: \textit{what parameters to learn} (which model parameters are important to prune the parameter space in a data-driven manner taking into account the dependencies like in~\cite{sophia-2019}, \textit{which model to learn} (the trade-off here is the usual bias vs. variance or we can take into account the model interpretability vs. expressiveness), and \textit{how to learn} (how much to generalize for example). 

One critical sector where model-based RL can come in useful is energy savings from electric grid-based applications. This falls under the ambit of automated decision making in industrial control systems. In this context, RL is more of a tool or decision assist rather than a scenario where the agent replaces the human-in-the-loop. 
Another relevant application of RL is in creating an optimal schedule of jobs on some physical resource pools (such as, machines on a factory floor). Some work has been done in the context of optimizing job scheduling within a factory model that will result in maximizing the rewards using ``approximate'' $Q$-learning where neural networks were used in approximating the $Q$-function~\cite{Minerva-2018}. 

\subsection{Domain-Specific Languages (DSL) for extensible CPS} Rather than vertical silos of CPS designed as a single-point solution, capabilities to support the next generation, extensible CPS will be needed. This can be helped by DSLs that aid in easily creating resilient CPS applications. 
There is only limited work on DSLs for CPS. One example is CHARIOT, which is a textual DSL developed using Xtext~\cite{thramboulidis2016uml4iot}. It provides an API that lets a developer create cyberphysical applications while remaining completely agnostic to the underlying middleware that can be used
by these applications to interact with each other. It also explicitly addresses resilience by allowing application designers to model systems' resilience
objectives. These become part of a configuration space that will be used at run-time to support self-reconfiguration mechanisms. 
There has been some initial work on creating DSLs for more specialized emerging application domains, \eg in the computational genomics area~\cite{sarvavid2016}. In the Sarvavid DSL, the authors identify recurrent kernels across the diverse swath of computational genomics and stitch them together using a domain-specific compiler that understands the dependence structures and domain-specific abstractions. These include loop fusions to prevent locality-sensitive cache misses, loop invariant code motion (to hoist a statement outside the loop body), and fine-grained parallelization~\cite{orion2014}, whereby Sarvavid can invoke an aggregation function on the results of all the partitions as Reduce tasks. 
Further, in the realm of IoMT, optimizing performance while respecting safety constraints will become critical, say in the context of a pacemaker for example. 

\subsection{Safety and performance guarantees for CPS} 
One of the major challenges associated with resilient CPS is the ability to reason about  
the system configurations associated with failures.  
Formal methods are an established tool to provide verification of desired properties (or of the failure to maintain those properties) in CPS.  These properties often are posed as logical constraints on the state-space, such as a need to avoid ``bad'' regions and to reach other ``good'' regions.  
The primary challenge in verification of CPS is in developing computationally efficient methods to determine if the desired properties can be met from a given configuration~\cite{flowstar, cora, Mitra, taliro}, or to identify the set of configurations which are assured to do so~\cite{Claire2003, chen}.  
However, unlike typical verification problems, which presume well defined, mathematical models of the CPS and any disturbances acting on it as well as clearly specified failures, protecting against failures in resilient CPS is complex. It requires being able to handle CPS with embedded learning elements, in environments with poorly specified disturbances, without attempting to enumerate failures ahead of time.

Computational challenges in verification of cyberphysical systems arise from the interaction of discrete and continuous elements, as well as from the existence of nondeterministic inputs, both controlled and uncontrolled.  The question CPS verification tools typically pose is, {\em Is it possible for any of the states within a given initial set to evolve into an \textit{a priori} ``bad'' region of the state space?}  
However, when inputs are probabilistic, it is more meaningful to focus on the likelihood with which desirable properties are assured, rather than a binary assessment of satisfaction.  
The framework of discrete-time stochastic hybrid systems captures rich behaviors associated with the integration of continuous and discrete processes and inputs~\cite{abate, summers}.  
Computational tools for discrete-time stochastic hybrid systems that avoid enumeration of state, input, and disturbance spaces exploit model structure, to either reformulate the verification problem as a convex optimization problem~\cite{sreachtools}, or to create an abstraction~\cite{faust, stochy, belta, kapinski} and employ simulation or falsification methods.  Controller synthesis with feedback largely remains an open challenge.  

One critical hurdle to extending existing frameworks to resilient CPS is that most methods do not scale well with the system's size.  Recent breakthroughs~\cite{hylaa, bak2019hscc, juliareach} have yielded results for non-stochastic systems with thousands to millions of continuous states, and demonstrated the need for tight integration with big-data approaches.  
However, many resilient CPS exceed this: consider the US power grid, for example, which, despite its known topology and dynamics, eludes formal modeling due to its sheer size and its dependence upon stochastic demand and generation phenomena. To make progress in formal assurances for resilience, significant advances in scalability must occur, which can handle the existence (and synthesis) of inputs.

As autonomous systems become ubiquitous, new approaches are needed to handle modeling challenges associated with neural nets~\cite{Taylor, pltl, verisig}, beyond simple, deterministic activation functions.  
Some data-driven approaches that do not require a mathematical model have been explored~\cite{dryvr, topcu}. However, significant work remains to fully understand the effect of sample bias and other phenomena related to finite data, as well to synthesize controllers in this framework.  
Advances are needed in the development of model-free or model-flexible methods, whose reliance upon data can be quantified with hard assurances. 
Additionally, another significant challenge in applying CPS verification frameworks to problems of resilience is that they require specification {\em a priori} of faults.  In many cases, such enumeration is simply not possible, due to the size of the system or even the inability to anticipate all ``bad'' configurations.   This is an aspect which, as methods for assured autonomy advance, may be ripe for integration with learning elements. 

\subsection{Distributed and dependable computing for large-scale CPS} 
One of the challenges of large-scale CPS design is that there is homogeneity in the code base and the operating system that executes on these devices. For example, 
there can be a large number of Programmable Logic Controllers (PLCs) in an Industrial Control System (ICS), all running the same software and wirelessly connected together. This opens up the possibility that a single exploited vulnerability can disrupt the entire CPS. This is reminiscent of the problem of monoculture in desktop OS  causing security concerns~\cite{goth2003addressing}.

Protecting embedded devices in the presence of vulnerabilities poses unique challenges that are fundamentally
different from desktop or server systems, simply porting defenses is therefore not an option.
First, embedded systems often run directly on the hardware without an intermediate operating system (or a thin OS layer) or a virtualization layer. The program itself is responsible for mediating access to all resources, including
security-critical ones, among all the tasks. Second, due to the lack of a Memory Management Unit (MMU),
embedded systems have a single flat address space where all memory locations (\eg the locations of I/O
ports) are static. Third, embedded systems are custom tailored to the deployment. In one specific hardware configuration, some I/O ports are security sensitive while others are
not. Orthogonally, even with significant advances, desktop defenses are incomplete and cannot defend against all code reuse
attacks or information leaks as shown through any recent attack that bypasses all existing defenses ~\cite{rudd2017address,schuster2015counterfeit}.
Leveraging buffer overflows, use-after-free bugs, indirect code pointers, or type confusion vulnerabilities,
adversaries can leak information and compromise software running on desktop systems despite all currently
used defenses.

There are several promising directions that have been developed to counter the above problem, in the context of CPS. They leverage the following characteristics of CPS, which provide some unique opportunities for strong, novel defenses. First, whole program analysis on these systems is feasible. Due to cost, power, and environmental constraints, systems are generally kept small. Best coding practices result in limited stack depth, restricted use of indirect control-flow, limited use of recursion, and fixed memory allocation. These bound the exploration space so that static analysis can be applied to entire programs. Second, the source code of each component is generally available to the developers as all components are developed by the same company. Even when libraries are used, all code is compiled to a monolithic binary and combined through Link Time Optimization (LTO). Third, both the software running on an embedded system and the underlying hardware are single purpose, further simplifying the analysis. The software running on embedded systems has limited functionality, often with a single purpose – compared to desktop systems with hundreds of parallel processes. Similarly for the hardware, each hardware unit is dedicated to a single purpose where on desktop systems the devices are shared among multiple processes. The combination of these opportunities enables us to scale static and dynamic analysis techniques to full embedded systems and to devise strong protection mechanisms that take these unique constraints into consideration.

Recent work to
protect bare metal embedded devices include  EPOXY~\cite{clements2017protecting} and ACES~\cite{clements2018aces}. EPOXY is an LLVM-based mitigation that enforces
a light privilege overlay, dropping privileges for all instructions and selectively raising privileges for a
few privileged operations such as writing to I/O registers. Based on this privilege overlay, EPOXY enforces
data execution prevention to prevent code injection, a safe stack~\cite{kuznetsov2014code} to protect against return oriented
programming, and diversification to protect against data-only attacks. ACES builds on EPOXY and automatically infers and enforces inter-component isolation on bare-metal systems, thus applying the principle of least privileges. ACES takes a developer-specified compartmentalization policy and then automatically creates an instrumented binary that isolates compartments at runtime, while handling the hardware limitations of bare-metal embedded devices.

Despite some existing solutions, there is still the need to develop security mechanisms for CPS against a wide variety of attacks, {\em without the need to re-architect the entire software}. This needs targeted static analysis to identify the
control and the data flow in the program. This can be augmented with an IoT emulator, which will
emulate the runtime behavior under fuzzed and controlled inputs to discover information not possible
through static analysis. This should be complemented with runtime enforcement techniques for enforcing
the principle of least-privilege execution, which is considered standard security practice, but is absent
in embedded system execution. Such a technique should be adjustable in the security-performance tradeoff space with feedback provided by the
constraints of the hardware and the performance impact due to the isolation of multiple compartments
of code and data.

\subsection{Generative platform development methodologies} 
In addition to the various specific challenges of CPS 
design outlined above, many of the broad and overarching challenges 
that CPS designers face can be traced to the need to
achieve flexibility and modularity on the one hand, and
high performance and resource efficiency on the other 
hand, \emph{at the same time}. 
Classical software and platform development methodologies are
not always able to resolve this tension, evidenced by the 
fact that resource-conscious developers tend to avoid 
adding abstraction layers in software systems for fear
of performance degradation. Despite decades of engineering,
compilers for general-purpose programming languages are
typically unable to optimize programs to a degree that would recover the indirection cost of such abstraction layers, leaving platform developers with the choice between
sub-par performance and overblown resource consumption
on one side, and essentially unmaintainable code that needs
to be painstakingly rewritten for each model generation 
on the other.

In computational genomics for example, the domain-specific language (DSL) Sarvavid~\cite{sarvavid2016} can express entire genomics applications in a few lines of code without requiring genomics researchers to be algorithmic or data structure experts. It uses kernels or functional building blocks for expressing genomics applications in an intuitive and efficient manner.
Overall, in high-performance computing (HPC), there
has been a shift in recent years away from low-level code
in general-purpose languages to DSLs
and generative programming~\cite{DBLP:journals/cacm/Ragan-KelleyASB18,DBLP:journals/pieee/FranchettiLPVSJ18,DBLP:journals/tecs/SujeethBLRCOO14,DBLP:conf/snapl/RompfBLSJAOSKDK15}. 
The underlying idea incurs a
radical rethinking of the role of high-level and low-level 
programming languages: instead of writing all the code in
a single language, the central idea is to use a high-level 
language for \emph{composing} pieces of low-level code, 
making domain-specific configuration and optimization a 
fundamental part of the system logic.
This direction holds a lot of promise for the world of CPS 
as well, as it fundamentally separates configuration of 
a system from its actual execution. Having a 
tightly-optimized execution path not only improves
performance but also reduces the surface for attacks,
and programmatic generation of an executable can be
combined with generation of contracts and proofs, which
has been shown to increase the scope for formal analysis 
and verification~\cite{amin2017type}.  


Together, the approaches described in this section show that our knowledge from a variety of fields can be leveraged to provide guiding principles for designing resilient CPS, and also for evaluating the resilience of our proposed designs. In particular, we have argued that a variety of approaches, ranging from following paradigms such as those of the fields of fog computing, multi-agent systems, game-theoretical modeling, dependable and secure computing, and domain-specific languages, to design more robust CPS, to the use of formal proofs to evaluate the security of ML-driven CPS and other safety and performance guarantees, to the use of reinforcement learning methods to simultaneously design and evaluate our CPS, all can present starting points for research into resilience-by-design.  

\section{Resilience by Reaction}
\label{sec:reaction}


\noindent In this section, we lay out some of the foundational technologies for resilience by reaction at runtime where the system is not designed with all the applications in mind but is agile enough to react to specific situations and then take steps to ward off the perturbation. This is because highly optimized tolerance~\cite{carlson-complexity} and specialized CPS may be more prone to failure against unexpected inputs. This is similar to how an ML algorithm can overfit to the training data, fitting tightly to the idiosyncracies of the training data, without being tested on unseen in-the-wild data. Plus, attackers can exploit the assumptions for which the over-optimization has been done. 
In addition to over-optimization, over-provisioning (\ie simply increasing CPS capacity to achieve  resilience) may make the impending failures even more catastrophic.   
For example, with the surge of smart (or, self-driving) cars, individuals who rarely drive will increasingly resort to cars, making the transportation situation worse, with unoccupied cars also ``traveling'' rather than parking---the commonly known Braess' paradox wherein building new roads increases congestion, pointing to the perils of over-optimization and over-provisioning. 

Another aspect that is not often considered is ways to prevent ML models from degrading over time, which calls for drift detection after model deployment. Model deployment should be considered the start of the model's lifecycle. These are some of the ways in which models can drift: 
\begin{itemize}
\item \textbf{Concept drift} originating from new ways of conducting fraudulent transactions for example. 
\item \textbf{Data drift}, where the data changes due to the changes in trends, seasonality or degrading sensors (that is how sensors need to be recalibrated over time). 
\item Finally, there could be \textbf{changes in upstream data} such as \textbf{changes in feature encoding} or changes in the type of (genomics) sequencing instrument generating the data, resulting in different error characteristics. 
\end{itemize}
Therefore, it is crucial to monitor the ML models after deployment.
We therefore alternatively explore solutions for resilience-by-reaction. 

\subsection{Serverless computing} 
There will be 12.3 billion mobile-connected devices by 2022, including machine-to-machine (M2M) modules, exceeding the world's projected population at that time (8 billion) by 1.5X. With the increasing evolution of IoT and edge computing, there is an over-abundance of geographically dispersed edge resources that are under-utilized. 

This is largely because of the inability of the cloud computing infrastructures to handle with high throughput the increasingly high volumes and varieties of workloads, with low latency desirable in many IoT scenarios that require real-time sensing, actuation, and dispensing such as in precision agrodispensing applications. In such scenarios, serverless computing and its pay-as-you-go model where the client only pays for resourced leased out by the cloud provider and with no unfront cost (elasticity) is ideal.
Thus, serverless computing is a cloud-computing-base execution model where the cloud provider manages resource allocation and the pricing model is based on the actual consumption of resources by the applications, rather than on pre-purchased allocations.

Plus, the ability to partition applications automatically, based on dynamic conditions, such as latency requirement, network bandwidth, other shared applications, and unanticipated circumstances, will improve resource allocations. Thus, instead of partitioning the different compute resources by design, the requirements will be partitioned---we call this {\em \bf top-down partitioning}---to generate the rules to drive the analytics at each of the three layers.

\begin{wrapfigure}{L}{0.60\textwidth}
 \centering
 \includegraphics[width=0.58\textwidth, keepaspectratio]{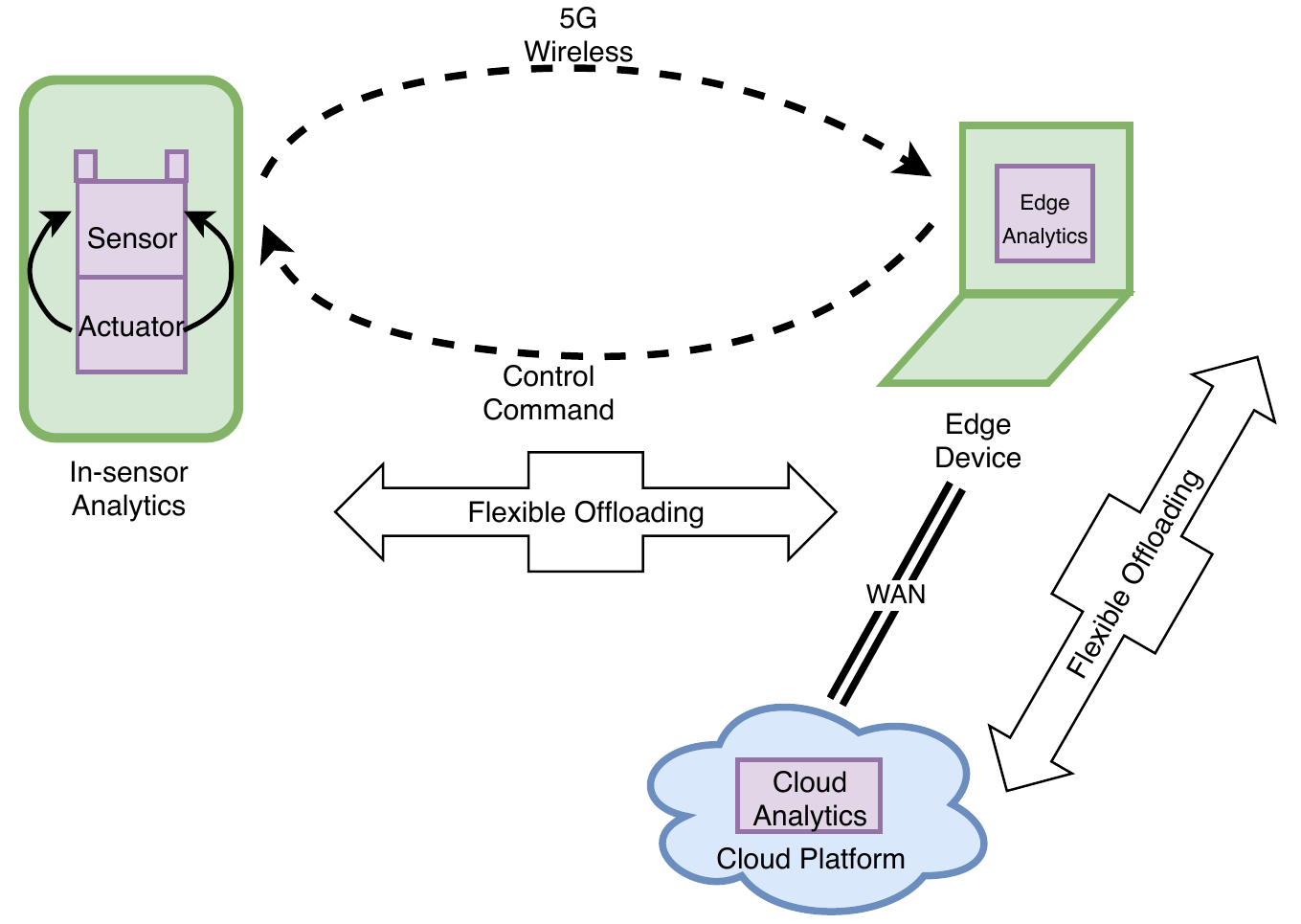}
 \vspace{-0.15in}
 \caption{Top-down partitioning for low latency.}
\label{fig:overview}
\vspace{-0.2in}
\end{wrapfigure}

Relevant to in-sensor analytics for example, the rules about when to compute the analytics answer and when to send results will be one of those rules that our top-down partitioning will identify.
This will improve the utilization of resources while reducing the cost. 
Amazon web services (AWS) is one of the most popular public cloud platforms. AWS-Lambda is what their serverless computing platform is called, and Lambda makes it possible to run code in response to specific demands while managing the compute resources available to the user.

\subsection{Top-down partitioning at the edge-cloud continuum}
Latency-sensitive applications will benefit heavily from the top-down partitioning of tasks across platforms, \eg IoT applications in digital agriculture [Figure~\ref{fig:overview}].
Ideally, with an eye toward optimal energy consumption, lightweight analytics would be performed at the level of the sensor nodes deployed on the farms (\eg ground sensors) or in the drones. The computational requirements at the level of the sensor nodes need to be lightweight in terms of the computational power, memory usage, and intermittent computation. In terms of the nodes' power consumption profiles, designers have at least three options available to extend the battery life and minimize the system maintenance of the sensor nodes: hardware optimization, software energy optimization, and energy harvesting. 
In terms of hardware, the primary energy consumption of a sensor node comes from its microcontroller, although the diodes, capacitors, and resistors have leakage currents. The software needs to be written to be event-driven and if identified as having a high energy consumption, the developer may need to optimize the speed rather than the code size.

\begin{wrapfigure}{r}{0.57\textwidth}
\begin{center}
  \includegraphics[width=0.55\textwidth, keepaspectratio]{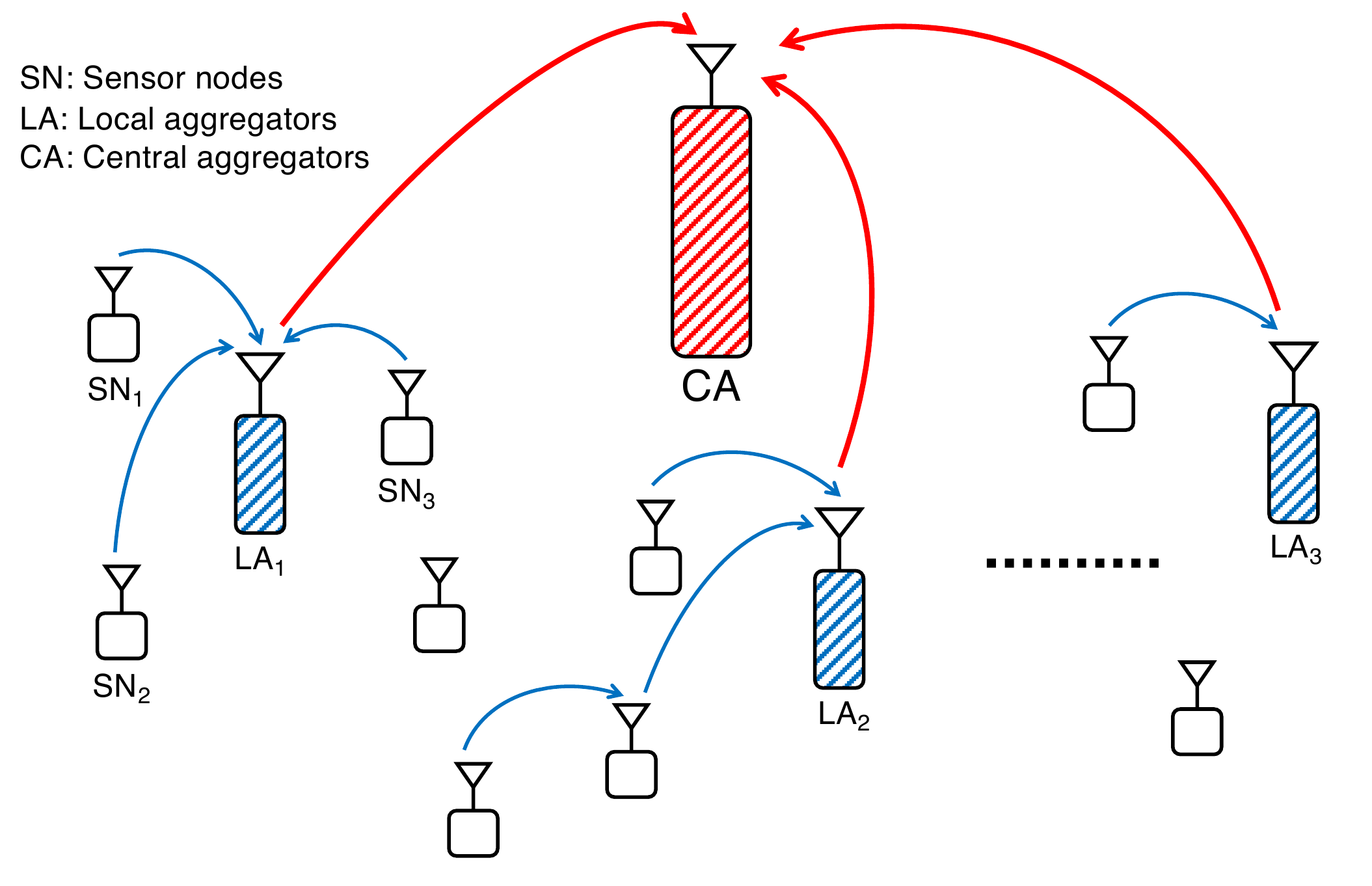}
\end{center}
\vspace{-0.3in}
\caption{Network architecture for scalability, balancing throughput with robustness and efficiency.}
\label{fig:network}
\vspace{-0.4in}
\end{wrapfigure}
Finally, there is a sizeable mass of work, \eg~\cite{energy-harvesting-2011, energy-collision-infocom-2018} on energy-harvesting sensor networks, which will be conducive to the digital agriculture or other IoT deployment settings and the detailed discussion of this literature is outside the ambit of this paper.

\subsection {Robust network operation} 
A hierarchical architecture, as depicted in Figure~\ref{fig:network}, balances costs, robustness, and reliability. 
The network nodes are divided into three categories: sensor nodes (SN), local aggregators (LA), and central aggregators (CA). This design borrows from an extensive understanding of prior work, \eg~\cite{ahmad2015survey},  in multi-hop networking, where smaller clients use low-power transmissions to reach local aggregators, then the local aggregators use higher-power transmissions to reach the central aggregators. 
Further, sensor-to-sensor communication mode is also available, which is crucial for network robustness.

In the context of optimizing the energy efficiency of the entire sensor network operation, one challenge in the digital agriculture scenario, for example, is the heterogeneity in the collection methodology (\eg mobile nodes in the form of tractors, trucks, or drones) \textit{and} the heterogeneous data sensors.
Here, differential spatial Nyquist frequencies become relevant to determine the collection frequency and the placement of the (static) ground sensors.

\subsection{Reactive resilience for mission-critical 5G}
Resilience in 5G networking can not be simply achieved by design. 
The lessons in 4G networks show that mobile applications may still suffer from bad performance, lost network access, reduced availability, and common failures and even
attacks when the built-in network capacity suffices. 

It is hard, if not impossible, for the network to function at its full potential under real-world complexity. This unfortunately thwarts resilience especially for those mission-critical applications like autonomous driving, tactile healthcare, and public safety. 
To tackle it, the emerging 5G system and applications should be able to learn what, why, and how regarding network behaviors under both normal and failure-prone cases and to adapt accordingly.
Rather than passively following the pre-defined network functions, they should proactively monitor network services and detect or predict whether any failure or misbehavior will occur with verifiable primitives. Given detected failures or performance degradation, they reason about the root cause and react in time to avoid or alleviate unnecessary losses. 

Moreover, reactive resilience can be accomplished at two stages: reactions under the existing infrastructure and reactions with evolved infrastructure. The former seeks a software solution that is allowed under the existing network hardware, mechanisms, and protocols. It is compatible to the deployed network infrastructure and seeks for opportunistic improvement. 
The latter takes the lessons gained in the previous network infrastructure and seeks for more profound changes that may upgrade network functions, protocols, and mechanisms. This matches with the legacy cellular network evolution 
where operators leverage flexibility in network management for short-term performance improvement over the current network technology (\eg 3GPP release 14 for 4G LTE advanced) while seeking for long-term enhancement by upgrading network technology to next generation (say, 3GPP release 15 for 5G).  

\subsection{Energy-aware computing} 
New paradigms, such as cloud and edge computing, in addition to the huge proliferation of IoT devices, enable billions of miniaturized devices to interconnect intelligently. Further, miniaturization of these devices is at risk because of the decaying Moore's Law and further miniaturization causing a shift in the performance and power-consumption behavior of otherwise-identical nanoscale circuits. Manufacturers try to deal with these huge performance and power variations by adopting overly conservative safety guarantees and redundant error-correction schemes. 
These pessimistic timing margins have severely degraded the electrical efficiency of computing. Reducing processors' energy consumption becomes even more important with the increasing numbers of interconnected devices, projected to be generating roughly 77 exabytes per month in 2022. One way to handle this continuous data surge is to avoid the pessimistic handling patterns for heterogeneous architectures and not artificially constraining the chips based on outliers. Instead, allowing them to operate at their true potentials. This will require embedding of diagnostic firmware-level daemons to reflect the altered operating potentials of the processors' and memory resources' capabilities even as these potentials change over time. Then, we could use optimization algorithms to maximize the overall throughput of the systems, similar to machine check architectures as in mcelog daemons in x86-based systems. For the actual optimization approaches, one approach would be using a combination of neural network-based approaches for building a surrogate model for learning the different performance metrics mapped to different workloads and node parameters, Then, rapidly searching through the parameter search space using genetic algorithms as in~\cite{rafiki-2017} and~\cite{sophia-2019} in the database optimization space for fast changing workloads.
Further, currently most of the processing of production-scale data sets takes places in the cloud, in massive centralized datacenters that are power hungry, deploying multitudes of servers and expensive cooling mechanisms. Decreasing the reliance on these massive datacenters and using the edge servers (with the surge of decentralized ``datacenters'') will allow pre-processing and selective forwarding of processed data sets to the cloud, not only making computing more efficient but also automatically improving data privacy because of the proximity of these servers to the client sites. Other techniques for energy optimization include optimization of DRAM refresh rates on the hardware side~\cite{DRAM-2019} and optimizing the mixing of low- and high-precision floating point operations for mixed precision settings using techniques as described recently in the GPUmixer~\cite{GPUmixer-2019}. 

\textbf{Algorithmic energy saving and model pruning mechanisms:}
Implementing large-scale deep neural networks with high computational complexity on ubiquitous and inexpensive IoT devices are invariably limited by the low compute capacities of these devices making it hard for these IoT devices to respond in real time. Pruning deep networks will enable their use on resource-constrained, battery-powered devices, such as smartphones and wearables, for running compute-expensive algorithms such as for object detection workloads. With this in mind, algorithms are being designed that use energy consumption of a CNN as a guide to the pruning process, reducing the energy consumption of large CNNs such as AlexNet and GoogLeNet by 3.7x and 1.6x, respectively~\cite{netadapt2018, yang2017designing}. Along the same lines of interpretability and efficiency of neural networks, physics-guided neural networks (\eg~\cite{jia2019physics}) also reduce the exploration space by introducing a second (physics-driven) loss function in the learning process, which relates to \textit{physical inconsistency} (in addition to the typical RMSE metric). Such reductions in the exploration space make the networks more generalizable, explainable, and efficient. Finally, it may be helpful to understand how regularization in neural networks is distinct from those in other areas of ML and quantitative tools such as the \textit{Weight Watcher tool}~\cite{martin2019traditional, martin2019heavy} can compare the accuracy of the pruned/regularized neural network with the original network. This is helpful because with two DNNs with the same (or similar) architecture, trained on the same data set, but trained with different solvers, hyperparameters, and regularizers, the Weight Watcher tool can predict the upper bound on the test accuracy without having access to the test data set. This is especially helpful where we try to deploy neural networks on embedded systems with limited hardware resources and the (Weight Watcher) tool tells us at what point further approximation/pruning of the network would break the model. 
Such quantitative metrics to be able to generalize neural networks, and in some cases, prune them without breaking the model is very relevant in today's CPS given the proliferation of neural network-powered devices.

Overall, the ideas presented in this section, regarding potential approaches to resilience-by-reaction, center around two main themes: a hierarchical architecture or top-down partitioning to isolate the effects of failures throughout the CPS, and using information from prior CPS deployments, or historical failure information, to inform future operation. We emphasize that along both these themes, and within the contexts described herein, there is a need to pay attention to both hardware and software solutions that can provide reactive resilience.   


\section{Application Drivers}\label{sec:drivers}
Our discussion so far has explored two complementary approaches to resilience: resilience-by-design, and resilience-by-reaction, which are largely applicable to any CPS. We elaborated on how our knowledge from a variety of fields of research can be leveraged to provide guiding principles for each approach to resilience. To further elaborate, we next focus on three specific application drivers. Considering the particular context in which the CPS for each of these application drivers will be implemented, as well as its design goals and limitations, allows us to further discuss application-specific research areas and challenges, especially from the standpoint of the application's scale and privacy requirements. 

The selection of the application drivers was with the intent to have a canonical (medium-scale) driver, covering both urban cities and rural (albeit, digitized) agriculture, then to move extreme-scale in the form of planet-scale where one has to think of scalability as a key feature in algorithmic design, and then finally concluding with wearables under IoMT. The latter brings with it a fresh set of challenges under the purview of privacy and human-centric features. These application drivers have humans built in-the-loop and this aspect is starting to gain traction in algorithm design in the form of humans-in-the-loop computing and HCI.


\subsection{Smart Cities and Digital Agriculture Systems}
\label{sec:smart-cities}

%
\textit{A smart city is a designation afforded to a city when it incorporates ICT to enhance the quality and performance of urban services such as transportation, utilities, and energy to optimize usage and costs.} 
Such a city also uses different IoT devices to collect data and draw actionable insights in order to optimize the use of limited resources. 
The many advances in information, computing, and communication systems, has led to increased interest in using these  technologies to help improve citizens' lives in urban and rural areas. These smart cities are envisioned to leverage the capabilities of ubiquitous IoT devices and CPS, together with the communication capabilities offered by next-generation networks (such as 5G), for data collection and processing, so as to enable real-time resource provisioning,  management, and planning at the city level. Advances in this domain can affect many areas of  operation within cities, including energy, transportation, communication, health, education, public safety, and  government. Effective design is based on algorithms and human behavior data rather than based on infrastructure alone. For example, peer-to-peer (p2p) services such as Uber and AirBnB make connections between the physical and the digital worlds without (or with limited) physical infrastructure such as physical offices. Another facet of a smart city will be the ability of a smart city to engage with the problems that are \textit{food-water-energy-nexus}-sensitive, and in this vein, reduction of pollution and greenhouse gas emissions can be a goal for smart cities. Understanding the challenges posed by the implementation of IoT and CPS technologies at the scale of smart cities, and for each of these sub-domains, presents exciting opportunities for interdisciplinary research. 
In addition, collective and correlated city living also degrade aspects of physical security related to pollution, disease propagation, and crime. 
The cyberphysical networks to be implemented in smart cities will improve various aspects of urban life but network vulnerabilities and cybersecurity challenges add new challenges. 
Here, we focus on these challenges from the perspective of ensuring resilience of IoT and CPS implementations in the context of smart cities. We also touch upon the emerging field of digital agriculture, in relevant parts of this section, and how a combination of emerging mainstream technologies and other specialized technologies will fuel the advances of digital agriculture systems.


\textbf{Components of a City-wide Cyberphysical System:}
A smart city deploys many different IoT devices to collect data and draw actionable insights for deploying within the framework of the city. Citizens expect at least five types of services from a city: namely, energy, transportation, communication, waste management, and (political) administration. Modern CPS networks will want to address these needs by variety of systems \eg smart energy  management involving residential use, street lighting;  smart environmental monitoring for air quality, sewerage disposal, and noise mitigation; smart transportation, including parking, traffic congestion control; and smart structural health monitoring involving buildings, roads, and bridges. These heterogeneous networks use a variety of protocols (\eg 3-5G, Wi-Fi, Bluetooth classic, or Bluetooth Low Energy (BLE)), update frequencies (for minutes to hours), power sources (main, battery, energy harvesting), and embedded security levels. Although a fully integrated network is planned for the future, the model smart cities today implement a subset of services.

\textbf{Current Status of Smart City Security:}
On the software side, the security of the smart-city service network is generally replicated and adapted from those in the traditional computer and communication network protocols. Similarly, the hardware reliability is based on a protocol similar to traditional integrated circuits. We will discuss the hardware aspects of the security challenges later in the report. However, the key challenge of ensuring the security of smart cities is the lack of a foundational understanding of the benefits and pitfalls of hierarchical and correlated networks and our ability to design adaptive security protocols that organically evolve with the network size. Existing smart cities differ in terms of capabilities and integration. Some prominent examples include cities such as Singapore and Bristol (United Kingdom),
which focus on government function and city services. In contrast, Google's \textit{Sidewalk Labs} being implemented in Toronto's Eastern Waterfront aims to tackle the challenges of urban growth while coming up with new standards of sustainability, affordability, mobility, and economic opportunity. Chicago's \textit{array-of-things (AoT)} is an urban measurement project with a network of interactive, modular devices or ``nodes'' installed in Chicago established as a ``fitness tracker'' for the city, measuring factors that affect livability in the city. Further, AoT has a rapidly expanding partnership program, mostly through university-based alliances, with nodes being installed in Seattle, Portland, Palo Alto, Denver, Syracuse, Chapel Hill, Atlanta, and Nashville. Even outside the U.S., there are devices to be installed in Bristol (England), Santo Domingo (Dominican Republic), Taichung City (Taiwan), and several Australian cities. 

\textbf{Desired End State:} We list below some of the desired properties of a resilient smart city such as having humans in the loop (when decision making models are faced with complex systems) and incorporating graceful degradation in IoT sensors, especially in the case of network dependencies. This list is not an all-inclusive list but highlight some key properties that can be considered in the evolution of smart cities.
\vspace{-6pt}
\begin{enumerate}
    \item Resilience against cascading failures by decentralized control, component-level heterogeneity (\eg Phasor Measurement Units (PMUs) in electrical power grids), and creation of resilient-state estimators.
    \item Develop metrics to quantify the degree of network correlation among inter-dependent networks.  For example, the failure of power grids (energy) could also affect self-driving cars (transportation).  Explore if solution in one space (\eg micro-grids may make power-systems more resilient) brings in other dependencies (\eg on communication network) that may actually make the overall system more vulnerable.  
    \item Encourage the use of context-aware sensor fusion. In other words, adopt orthogonal, multi-probe sensing for command and control. For example, a corrupted central monitor may be fed incorrect data from smart air conditioners so as to increase the power delivery abruptly, leading to a cascading failure. An independent sensing of the local weather, which shows no correlated increase in ambient temperature, may suggest that the original request is spurious and can be ignored. 
    \item Quantify the reliability requirements for various levels of the hierarchy because all failures are not equal. A bug in the self-driving car may compromise its performance, but a sudden shut-down/reboot of the system will be catastrophic. Develop hardware and software reliability protocols for these levels. Use CPS checkpointing and recovery/verification to handle inescapable failures. 
    \item Develop a network algorithm to identify graceful degradation of IoT sensors connected to the network. 
    \item Be cognizant of the pitfalls of over-optimization and over-provisioning. This includes overbuilding a city after a natural or man-made disaster.
    \item Repurpose resources, especially for resource-constrained rural areas. For example, drones in rural farms, which serve as low-cost aerial camera platforms, can also act as data ferries in these low-bandwidth areas.
    \item For critical networks such as bridges and the medical system, design the system with human-in-the-loop (HITL) considerations. These HITL considerations can be introduced in different scenarios such as: humans labeling data, humans tuning the model teaching the model edge cases or new categories in the model's purview, and testing and validating a model by scoring its outputs, especially when an algorithm does not have high confidence about a decision or is overly confident and is outputting incorrect decisions, think of biases in human judgements but instead in a model this time. 
    \item Create protocols for security updates, response to security incidents, and ensuring privacy. This has to be done being cognizant of scale and the fact that a smart city operations team will be understaffed and not have deep expertise in all aspects of security or resilience. 
\end{enumerate}



\textbf{Considerations in the design and integration of CPS resilience in smart cities:} In order to achieve the aforementioned end results at the scale of smart cities, there is a need to focus on building reliable systems and quantifying results in terms of specific metrics, instead of developing proofs for small-scale model systems. Throughout these efforts, researchers should use system-scale modeling tools to categorize the security needs of various layers of the hierarchy and work with hardware/software manufacturers to ensure adoption of tier-specific reliability protocols. An important consideration is to build reusable building blocks or ``kernels'', instead of building algorithms \textit{de novo}, to address these scalability and security challenges, as has been instantiated in the computational genomics domain~\cite{sarvavid2016}. Lastly, it is desirable to build algorithms with humans-in-the-loop (HITL), with the goal to ultimately move to humans-on-the-loop and greater autonomy. In fact, developing models with HITL (interpretability) decreases the number of users required in the study and incorporates interpretability into the training objectives, as in~\cite{interpretability-2018nips}. We enumerate below some of the considerations for integrating CPS resilience into smart cities.

\begin{enumerate}
\item{Resilience against cascading failures:}  
One of the first resilience challenges that emerges with the implementations of  CPS at smart city levels is the increased risk of cascading failures. These types of failures arise due to the dependencies among a large number of CPS devices, and also due to the interdependence between different CPS, such as between medical services and transportation networks, at the city level. Making CPS resilient against these cascading  failures calls for research in both resilience-by-design and resilience-by-reaction. 

\item{Resilience within each CPS:} 
One of the approaches to improving resilience of complex systems, proposed in  works such as ``The architecture of complexity", by Herbert Simon~\cite{complexity-architecture}, is to follow a hierarchical design and operation. 
However, we should also understand that this type of design can bring unanticipated cascading failures to these systems. In particular, CPS intertwine the operation of physical and software components. As such, failures in physical and cyber resources become interdependent as well. For instance, a physical failure of a control center, such as due to natural disasters, can impact the cyber components throughout the CPS, and vice versa, the failures of cyber controls can lead to damage of the physical resources. Our design of resilient CPS should account for the cascading failures that cross over the cyber and physical realms. 
In addition, a better understanding of the effects of resource heterogeneity (or lack thereof) is important to the operation of large-scale CPS. 
For instance, if a large subset of IoT devices comprising of a CPS utilizes the same form of software, a single software vulnerability can disrupt the CPS. This motivates research on the optimal design of heterogeneous CPS under cost and operation constraints. 

\item{Resilience across different CPS:}
More broadly, making devices smarter increases the connectivity between different CPS in smart cities, and will further increase the risks of cascading failures across CPS. This raises challenges on understanding whether, and how, different networks in smart cities can affect resilience?

\item{One-hop dependency:} For instance, within the smart city, routing of medical emergency vehicles that connect patients to healthcare providers can be facilitated using a smart transportation network. This type of service interlinks the two networks, making the resilience of the centralized emergency healthcare system dependent on that of transportation networks. 

\textit{Multi-hop dependency:} Cascading failures can spread beyond one-hop dependencies as well. As an example, consider an electric vehicle (EV) within the smart city; the operation of this vehicle is subject to resilience of both the electric grid and the transportation system. A failure in the electric grid will affect the travel patterns of EVs on roads, changing the expected operation of the transportation networks. This could in turn affect the routing of medical vehicles and the delivery of emergency healthcare. 

It is therefore important to make interconnected CPS within smart cities resilient-by-design against these types of failures. It is also of interest to understand how to reactively adjust inter-CPS dependencies and operations to block such cascades of failures. 

\item{Enhancing the availability of physical resources:} 
A potential method toward improving the resilience of CPS against either natural or malicious perturbations is to  increase the capacity and availability of both cyber and physical resources. Increased capacity will make the system resilient against failures due to resource scarcity, while increased resource availability can allow the CPS designers to build redundancies within the system so as to enable fast recovery following failures. A well-known concept in this regard is that of consistency level (CL) and replication factor (RF). A replication strategy determines how many nodes have replicas placed in them. That is, RF is equivalent to the number of nodes where data (rows and partitions) are replicated where data is replicated to multiple (RF=N) nodes. So, an RF of 1 means that there is only one copy of each row in the cluster. If the node containing the row goes down, the row cannot be retrieved. RF of 2 means two copies of each row, where each copy is on a different node. While such replication strategies are more or less routine for the cyber side~\cite{rafiki-2017, sophia-2019}, innovative methods are required to increase physical resources' capacity and availability. Improving resilience in this way falls within the category of resilience-by-design, but can also be done in a reactive manner. 

This type of design is of particular importance in smart rural communities for applications such as agricultural product yield (\eg \textbf{digital farms}), in which natural perturbations such as wildfires, flooding, etc. might eliminate the limited resources the CPS has to work with. The technical challenges are not necessarily the first challenges of resilience in these situations, rather resource availability has to be addressed first.

One of the methods for increasing the availability of physical resources is  \emph{repurposing} existing physical resources. For instance, drones used for aerial surveillance and data collection can also be used for data transportation or internet connectivity in rural areas. More broadly, idle resources on edge devices (\eg laptops, cameras, cell-phones) can be used in a similar way to replace traditional base stations or routers. 
For smart grids, microgrids and incorporation of new storage technologies can be used for adding capacity.
\end{enumerate}

\textbf{Digital agriculture-specific considerations:} In the context of digital farms, some of the critical bottlenecks are lack of connectivity and the need for connected sensors, connected opportunistically using the edge and sensor nodes to perform basic computations (\eg anomaly detection). For processing on resource-constrained devices, a tradeoff is needed between the computational load and accuracy, which may need approximate computing techniques. An example may be that the approximation of the output translates to the sensor relaying whether the soil moisture content is below or above a certain threshold rather than outputting the exact value of the moisture content. Further, the spacing of the sensors itself may be guided by information theory principles, such as the Nyquist-Shannon sampling techniques, which will also determine the required redundancy in the model for robust sampling from these sensors. Another consideration is the low-power communication technologies for wireless IoT communication, which broadly falls in three classes. Some of the bottlenecks in this wireless transmission include harsh farm conditions and the abundance of inexpensive and less reliable sensors plus the intrinsic challenges of implementing 5G networks and LoRa and LoRaWAN. For the latter, the challenges include data security, for which data transmitting technologies may have to avoid big name data servers such as MQTT, LWM2M, and CoAP. The latter may make clients more wary of usage. Also, there are competitors in this space such as SigFox and NB-IoT, with the latter requiring lower investment upfront. Finally, due to the license-free operating conditions of LoRaWAN, it operates at open frequencies that may incur interference, deterring the transmission of large data payloads. Following are some of the low-power wireless communication technologies in use:
\begin{itemize}
   \item Low-power wide area networks (LPWAN), with a greater than 1 kilometer range, essentially low-power versions of cellular networks, with each ``node'' covering thousands of end devices. Examples include LoRaWAN, Sigfox, DASH7, and weightless.
  \item Wireless personal area networks (WPAN), which  typically range from 10 meters to few hundred meters. This category includes Bluetooth and Bluetooth Low Energy (BLE), ANT, and ZigBee, which are applicable directly in short-range personal area networks, body area networks, or if organized in a mesh topology and with higher transmit power, larger areas of coverage are possible.
  \item Cellular solution of IoT, including any protocol that is reliant on the cellular connection.
\end{itemize}

\subsection{Planet-scale IoT}
\label{sec:planet-scale}

\noindent The International Data Corporation (IDC) expects that by 2025 there will be 41.6 billion connected IoT devices, generating more than 79 zettabytes (ZB) of data~\cite{IDC-projections}. This upsurge of devices alongside the evolving technologies, ranging from small (\eg energy-independent IoT devices) to large (\eg nature-inspired neural networks~\cite{neuroevolution-2019}), are beginning to adapt to technological shifts (\eg end of Moore's Law), resulting in new challenges in resilience, especially from the standpoints of energy efficiency, security, and privacy.
IoT has the potential to network an enormous range of connected devices---including home appliances and utilities, wearables, homes, corporate buildings, industrial processes, medical devices, law-enforcement devices, military equipment, and other futuristic applications whose impacts are hard to imagine in today's context. 
Some of those ``things'' will be directly accessed through the internet while others will be concealed in local networks behind firewalls and (network) address-translation (NAT) routers.
The goal of a planet-scale IoT is creation of a highly-available, geographically distributed, heterogeneous large-scale IoT-based system that has the same efficiency, maintainability, and usability as today's data centers. This has to be achieved in the face of a high degree of heterogeneity of the constituent devices, in properties such as computational power, energy availability, and resilience to errors and attacks. Further, critical technology drivers for IoT success will be AI, edge computing and 5G, digital twins, and blockchain~\cite{blockchain2019panel}.

Planet-scale IoT can be thought of as smart cities-at-extreme-scale. For this to happen efficiently, one has to think along the following lines: ``provably resilient'' energy-saving devices (ultra-low power and energy-independent, mitigating the flaws of outlier devices); self-learning technologies (\eg using reinforcement learning (RL) methods); ``provably'' secure artificial intelligence/machine learning (AI/ML) methods, both from the hardware and software standpoint, and able to tackle extreme-scale complexity based on formal methods, new abstractions, and other complexity management solutions. In terms of hardware scalability, one can think of vertical and horizontal scaling; \textit{vertical scaling} means adding to the number of cores of an existing machine, so more power in the form of CPU and RAM and \textit{horizontal scaling} means adding more machines to the pool of resources. Another related aspect is porting to the cloud or edge for ``horizontal scaling'' and this is where smart provisioning of instance types and sizes will come in handy as will the proper tuning  of the underlying database backends~\cite{sophia-2019}. Of course, ``scaling up'' resources is also possible on the cloud where one just adds more RAM, CPU, disk space, and network throughput, without making any modifications to the application itself, only changing the instance type and size. However, there are limitations to this vertical scaling up to the point where the application itself becomes the bottleneck. With the ``genomical'' data sizes, often noSQL backends are deployed to scale with the increasing sizes and diversity of queries, such as in genomics~\cite{scalable-2017-chaterji} and IoT-related domains~\cite{IoT-2017-chaterji}.  
Further, we have to keep in mind guarantees of safety, availability, and timeliness, either deterministically or stochastically in a complex environment, with humans-in-the-loop. From the standpoint of usability, given the rise of heterogeneity and specialization at multiple scales, domain-specific languages (DSLs) may enable some level of domain-level ease-of-use, stemming from modularity and reuse~\cite{sarvavid2016} or even make the choice of parameters for algorithm tuning (\eg deep learning algorithms) at extreme scale more manageable~\cite{demystifying-2019}. However, the rise of DSLs, along with the use of different abstractions, tools, and approximations by different domain scientists, even in quite limited domains, may make interoperability unlikely. To enable a functional IoT, a balance must be struck between modular, general-purpose, high-level languages and DSLs. An example of such a middle ground is REST (REpresentational STate Transfer) APIs for web services, which can serve as the go-between for software developers and users. In addition, federated computational infrastructures, as envisioned and under development for genomics applications will result in savings both from the computational and storage standpoints~\cite{federation}, albeit, may need to be instantiated in cautious steps, \eg in production-grade, heavily used environments~\cite{mgrast-2017}.

\noindent Next, we enumerate a snapshot of technologies that will enable resilience at planet-scale and some of the challenges arising from deployment at planet-scale, although the more beefy technological piece from a foundational perspective is later on in the article:
\begin{itemize}
\item What challenges are brought about by the large numbers of devices and that too heterogeneous devices? 
\item What challenges are brought about by the lack of perfect connectivity of all these devices on the planet? 
\item What challenges are brought about by obsolescence of devices across the planet, such as, legacy software and software upgrades, maintenance, recycling? 
\end{itemize}

\noindent \textbf{Network connectivity, sharing economy, and digital surveillance:}
\noindent While smart-mobility technologies exploit multi-modal mobility options, they can also augment the transportation and mobility planning in cities. Services such as Red Ride (US), Matchrider and Flinc (Germany), Lyft (US), and Uber (US) enable the userbase to find the closest rides available. Such mobility technologies enable sharing new data, generating new data, and importantly, involve citizens. For example, tagging in OpenStreetMap (\url{www.openstreetmap.org}) lets users tag geo-coded properties with features of interest, \eg BBBike Initiative in Berlin contributes to the ``bikeability'' index. The challenges arising from such crowdsourcing activities include: \textit{data ownership}, \textit{data privacy}, consistency, and accuracy. 
Growing complexity of car-to-car and car-to-infrastructure communication allow over-the-air wireless communications (for updates), online diagnosis, and maintenance, in addition to data collection opportunities for smart city and infrastructure planning~\cite{augmented-2019}. Learning from these smart-mobility technologies, while 5G and other forms of innovative network connectivity require new technologies, they also enable new uses. From the standpoint of novel networking technologies, design of adaptive routing algorithms that will organize the traffic automatically based on the availability of nodes will result in more efficient networking. Since sensor nodes are often static, designing light-weight routing protocols is useful. Further, network protocols that can relay the level of ``stress'' in terms of resource availability will result in alleviated network congestion, especially at planet-scale. Further, scalable estimation of network congestion without conveying raw information will be helpful

These also extend to digital farms (think digital agriculture~\cite{farmbeats-2017}) beyond the ambit of a ``typical city'', where in addition there are other shortcomings such as lack of wireless connectivity, exposing them to additional technical challenges and opportunities. To further connect these smaller farms more globally, think smallholder agriculture producing 80\% of the world's food connected using a farmer-to-farmer digital network (so in a way, planet-scale), farm connectivity and farm machine intelligence will go a long way.
Further, for low-latency operations either in the smart-city or digital farm setting where decisions have to be made on the fly (\eg actuating the dispensers of an agro-chemical machine as it ``senses'' the soil at real-time), effective partitioning of analytics between the sensor, edge, and cloud platforms will be critical. Such partitioning needs to happen in a combination of top-down and bottom-up manner. Top-down means that we take the high-level user requirements on latency and accuracy and define the partitioning based on that. Bottom-up means that depending on the available resources on each platform, we decides where to run the application component. Top-down requirements naturally have a higher priority.
While \textit{data ownership} is a big bottleneck in digital agriculture for which some companies (\eg Farmobile, \textit{AGData Transparent Certified}) are already compensating farmers for their Electronic Field Records (EFRs), \textit{data privacy} is a looming concern in digital health, which hinges on patients feeling comfortable sharing their most intimate data (personal health information, PHI), especially when such constant health surveillance data could affect their employment.

\subsection{Resilience in the Internet-of-Medical-Things (IoMT)}
\label{sec:IoMT}





Fueled by the recent proliferation of wearable health and activity monitoring systems, IoMT stands to revolutionize healthcare.  Wearables and implantable medical devices  are IoT devices that monitor the vital signs of a person continuously and wirelessly over a long period of time. Wearables have a long history of applications in hospital settings for continuous monitoring of vital signs (\eg heart rate, respiratory rate, pulse oximetry).
The security and resilience of the IoMT faces many challenges.
These challenges include \textit{confidentiality} (for sensitive information in the system), \textit{data integrity} (ensuring that the data in the systems has not been accidentally or maliciously deleted), \textit{software integrity} for the on-sensor algorithms running on the system, and \textit{availability} that simply means that the system is available to authorized users seamlessly. Availability also translates to safety, ensuring that denial-of-service attacks do not succeed. 

There are several attributes of the current state of IoMT devices and systems that present obstacles to addressing the security and resilience challenges.  Redundancy in hardware and software is a hallmark of most security and resilience solutions.  However, most IoMT platforms are siloed with limited APIs.  Commercially available systems (\eg Medtronic, Philips, Apple, Fitbit, Google, Samsung) are not designed to communicate with  one another and offer limited ability to access data in real-time.  Complicating the matter, data formats in IoMT devices and platforms utilize proprietary data representations that are not necessarily consistent with existing standards (\eg FHIR, HL7).  From a privacy perspective, in many IoMT devices, data privacy is subject to terms-of-use and provides little (if any) assurance of privacy when integrated with multiple sensors and platforms. Lastly, there has been limited prospective clinical use of many IoMT devices due to statistical variability in the data, with an exception being the FDA-approved Apple watch, although the data requires post-processing for clinical use.

Overcoming these privacy, security, and resilience obstacles requires improving the current IoMT.  First, to increase the redundancy in the IoMT requires the development of medical device interoperability platforms that support traditional medical devices and next-generation wearables.  Toward enabling medical device interoperability, there is a need for a categorization and standardization of the widely varying communication and security protocols being deployed by various healthcare companies. Developing platforms that are necessarily inclusive will promote manufacturer participation in resilient wearable platforms. At the same time, systematic methods and libraries to support patient data security and privacy need to be standardized.  

To bridge the gap between the current state-of-the-art and the desired end-state presents multiple opportunities.
First, developing a methodology to assign security-levels to various IoMT technologies depending on the consequence of the false-positive and false-negative events, the number of people affected, and the severity of the medical problem,  among others. 
Second, integrating and promoting platforms to support medical device interoperability that provide incentives for manufacturers to participate.
Third, creating tutorial materials that are easily understood by the device designers, including consolidating standards and libraries that can be integrated during the device design phase.
Lastly, developing application programming techniques that incorporate security and resiliency as a service, consequently, abstracting away much of the resiliency effort from clinicians and data users.

\section{The Road Ahead: Snapshot of technologies and learning from the cyberphysical resilience of living cells}
\label{sec:road-ahead}

Here we look at the road ahead with an appetite for intelligent and federated learning (distributed computation such as in blockchains) and also learning from nature's computation, quite literally, from the computation embedded in living cells. While learning the code of life will be instrumental in understanding and even designing synthetic life, there are resilience lessons to be learned from life's computation. Another aspect that we do not touch upon in depth but nevertheless want to briefly discuss here is the result from the maturation of some of these technologies at the frontier of the cyber- and the physical. Take for example connected and autonomous vehicles. As these autonomous vehicles become more ubiquitous, there would likely be a level of artificial intelligence imbued into these vehicles making them ``sentient'' or even ``anthropomorphic'', that is, human-like, as imagined by some early movies that anthropomorphized a heroic car in ``Herbie the Love Bug (1968)'', a sentient Volkswagen Beetle that rescued its owners from various predicaments. More recently, of course, there has been ``Cars'' (2006 onward) imbuing cars with personalities. All of this seems a little on the brink of tottering optimism and has resulted in the creation of the ``Future of Life Institute'' dealing with edgy subjects such as the ``emotional intelligence'' of assistive robots with the emergence of fields such as human-robot interaction (HRI)~\cite{assistive-robot-cdc-2018}.



\noindent \textbf{Using intelligent edge for low-latency IoT analytics:} 
The next wave of IoT innovations will be fueled by data science and engineering. While there is a plethora of cloud-based, centralized data analytics platforms (\eg Amazon Web Services (AWS) IoT, IBM Bluemix, and Microsoft Azure IoT Suite), cloud-based solutions suffer from limitations such as high-bandwidth cost and high-latency operations and are constrained by the assumption that the connectivity between these IoT devices and the cloud is sufficient for the continuous relay of information. Further, a vast swath of IoT devices may also be unsuitable for cloud connectivity because of regulatory, security, and privacy concerns. Such limitations are now partially obliterated by the increasing sophistication of IoT devices such that these devices have increased computing power and/or storage capabilities to be able to perform some of the analytics at the gateway- or device-level itself~\cite{sheth2016internet}. In addition, \textit{fog computing} (a term coined by Cisco), which could include devices such as Raspberry Pi, devices with ContikiOS/TinyOS, mobile devices with Android or TinyOS, or routers with compute capabilities, can dynamically adapt to context and application requirements rather than providing simple static services. This is where fog intelligence comes into the stratosphere. In addition, mist computing (again, coined by Cisco) pushes the computation to the extreme edge of the IoT environment, where a ``mist node'' (unlike a more rigid preconfigured sensor node) can provide simple application-specific tasks, \eg data integration and thresholding, data filtering based on the thresholds, or data cleaning. Another enabling strategy of edge computing is fine-grained parallelization as in prior work~\cite{orion2014, sarvavid2016, mahadik2019scalable} where the parallelization ability segments the tasks into smaller ones. These partitions can then be
offloaded into different fog nodes with different capabilities. The objectives of this kind of optimization could be a combination of the following: energy or bandwidth-saving capabilities, low-latency operations for specific use cases such as real-time video surveillance, smart city-based applications, or robotic surgeries. Further, fog nodes need to be adaptive, for example the locally processed analytics could be offloaded to the cloud to service remote users or is \textit{not} offloaded to the cloud for local users. In the latter case, localized protocols (\eg Wi-Fi, Bluetooth) are used for local dissemination. 

From the design considerations angle, here are some things to consider: use of ``surrogate fog nodes'' that can be used for adaptive offloading among the different fog nodes---the initiator node and other surrogates---based on the periodic monitoring of the status of the fog nodes (\eg availability, battery level, CPU, RAM, bandwidth, available memory, etc.). The array of ``features'' of the surrogate fog nodes will enable the intelligent offloading of the tasks to the most capable node. For such kinds of context-aware partitioning, some of these applications may need partitioning at the design time, or in concert with runtime partitioning, such that the user-facing parts of the applications are performed locally while the more compute-intensive applications are performed on well-resourced servers (so, remote execution). For further granularity in such kinds of partitioning tasks, docker containers rather than VMs may be better positioned because containers do not bundle the full OS, rather they bundle a docker image that contains the necessary executables and libraries required to run the tasks on the partitions of the fog node without over-taxing the smaller compute resources of the fog nodes. 

\noindent Some of the constraints of these intelligent fog environments in the context of smart cities are as follows:
\vspace{-5pt}
\begin{itemize}
    \item Heterogeneity of the fog nodes limits the use of existing data processing frameworks such as Apache Storm, which are more adept at handling static configurations. In such cases, newer frameworks, such as Apache Flink, which essentially handle stateful stream processing, may be closer to what we need.
    \item Current container orchestrations are primarily targeted at more homogeneous cloud instances and therefore efforts such as the instantiation of microclouds may be needed to bring container technologies to the IoT/edge devices.
\end{itemize}
Challenges arising from a heterogeneous cloud-edge continuum come from the heterogeneous protocols and hardware operating at different specifications. To address this, a common strategy is to have abstractions or middleware on the fog computing layer to allow microservices to run on fog nodes, \eg Eclipse's ioFog \cite{iofog}. While containers and VMs could also fulfill the above role, they have been built with cloud orchestration in mind, where heterogeneity is magnitudes lower and tasks are more resource-intensive.


\bigskip

\noindent \textbf{Privacy preservation through blockchain and decentralized federation:} 
Blockchain, the technology underpinning cryptocurrency, such as, Bitcoin and Ethereum, is a state-of-the-art, cryptographic, distributed \textit{and} decentralized ledger, with the potential to remove the middlemen from all transactions via the use of validators or ``miners".  The blockchain ledger works by permanently recording, in chronological blocks, the history of all transactions that occur between peers in the network. All validated transactions are recorded in the form of a chain of blocks, hence the name ``blockchain". In terms of which chains are the most trusted, proof-of-work protocols or functions create a \textit{distributed, trustless consensus}. The result is that the longest chain has the network's consensus and is the correct one and forks in the chain are discarded. 
Blockchain technology is potentially helpful for identity and access management (IAM) in federated infrastructures. By combining cryptographic hashing with blockchain technology, one can bring one's own identity, \ie the BYOI paradigm, to the federated platform and can receive the proper authentication. For example, the cryptographic SHA256 hash computational algorithm would return a 64-digit, fixed-length set of numbers called a hexadecimal signature for an input transaction (say, for personal genomic records of an individual or metagenomic signature of one's gut flora). Thus, changing even a single attribute in the input would create a distinct and random hash value, however, the same set of records would transform into the exact same hash value. By altering the hash of the input transaction, the access credentials of this input will change, allowing for different access credentials, for example. On the algorithmic side, Hash values are in turn combined into \textit{Merkle trees}---the simplest one being the binary Merkle tree, with more complex Merkle trees being possible such as the ``Merkle Patricia tree"~\cite{gonnet1992new} being used in Ethereum. 
The block's header now contains all of these hashing results, along with a hash of the previous block's header. This time-stamped header then becomes a part of a cryptographic puzzle solved by manipulating a number called the nonce. Once this puzzle is solved by miners (who then get a small reward), this new validated block is added to the blockchain. 

For example, for every human, a hash of all the personal information can be created and the personal information need not be stored. Then, when the time comes to check the identity of an individual, the identity can be checked against the archived hash without the validator (``miner") needing a copy of the original digital information deck.  This is going to come in handy as more and more genomics and metagenomics data sets pertaining to an individual are generated, including being amassed by the patient herself~\cite{kish2015unpatients} and the associated digital liability of these federated cyber infrastructures. There is increasing trend toward the use of cloud infrastructure for genomic analyses and a slowly emerging trend of protecting such computation with the use of privacy-preserving transforms such as cryptography and obfuscation. Now, with the maturation of the blockchain technology, it may well serve as the new organizing paradigm for healthcare-related data and IoMT, especially as the volumes and frameworks for such ecosystems abound~\cite{federation}. 


\bigskip

\noindent \textbf{Resilience of data sharing repositories and the emergence of BioIoT:}
There are over 900 environmental observation laboratories worldwide, having membership in data-sharing consortia such as the International Long-term Ecological Research Network (ILTER)~\cite{ilter2018}, EarthCube~\cite{earthcube2017}, National Ecological Observatory Network (NEON)~\cite{NEON2015}, TERN~\cite{TERN2013}, ICOS~\cite{ICOS2016}, and Natura 2000~\cite{natura2010}, etc. The common theme of sharing data, carrying out data aggregation, processing and analysis in web-accessible open repositories puts these largely ecology-based projects in the realm of the Bio-IoT. Collections of sensors are deployed at a particular location for which interesting scientific questions have been formulated. Some of the sites have been designed specifically to use modern remote sensing, spatially distributed clusters of instruments, and data collection, aggregation, processing and centralized access to repositories and models in order to assist scientists interested in biodiversity, ecosystem integrity, ecological forecasting, and understanding the forces that lead to ecological change. The scientific communities engaged in monitoring earth systems (polar, atmospheric, abiotic and biodiversity, ecological processes of energy and material flux, etc.) have developed standards and protocols that enable data sharing, providing harmonized measurements, automated quality assurance and data integration workflows. More recently there has been an initiative across the ILTER sites to provide a complete suite of sensors to make a network of sites of the networks. A sophisticated Data Science Architecture design for such projects can be identified in most recent projects, for example, the EarthCube cyberinfrastructure program, a community-driven project to develop information- and tool-sharing frameworks for geoscience-related discoveries. 
The notion of resilience at these sites has mostly focused on ensuring data capture, integration and quality to create actionable knowledge for specific constituencies, \eg researchers and local or national government agencies. Looking forward, as networks of information collection systems and the resulting knowledge become more important to decision making and value-based decisions, the systems will also become targets for interference, whether by shutting down the data collection or contaminating the data with false observations. However, resilience by design for any part of these systems, with respect to CPS as discussed herein, is rarely considered, as described in~\cite{EVO2016}. Resilience by learning is implied for some components, where sensor development (more sophisticated, robust or innovative) and deployment (in terms of increased coverage with respect to more sites or more sensors) advances can be taken as an example, but the expansion comes from scientists with narrowly focused interests rather than for provisioning heterogeneous collection of devices supported by citizens with less sophisticated but broader-spectrum capabilities and quite distinct interests in data use. Another aspect of resilience that has been discussed for these platforms and networks has to do with the Big Data applications that can arise if proper integration and quality assurance and accessible sharing resources are available. A common concern that has been expressed at this level is how to design alternate entry points and also provide for learning new entry points to make levels of data and models available that are appropriate to the needs of the user. 
Data wrangling is only important to a small technically sophisticated subset of users and so long as vetted workflows on properly quality assured data sets are guaranteed, the summaries and reports are better starting points for many users, whether those are in government agencies, NGOs, or small groups interested in local data, like farmers.  Thus, there are multiple opportunities for interdisciplinary projects that take on the CPS of large environmental virtual observatories, situated across the globe. 
\bigskip

\noindent \textbf{Learning from the regulatory network of living cells:}  
In many ways, epigenomic regulation offers ``CPS resilience'' to living systems. The term ``epigenomic'' mainly refers to histone modifications and DNA methylation, which are alternate ways to alter the gene expression landscape of an organism's cells in spite of having an identical (more permanent and static) DNA sequence. Other less overt means of epigenomic regulation are through genomic enhancers and through small regulatory RNA such as microRNAs (miRNA), which post-transcriptionally modify gene expression levels~\cite{tiresias-2018}. Overall, the genome is armored with multiple levels of redundancies to make it fault tolerant, masking it against the processing of incorrect signals, whether internal or external, such as might arise from a viral infection, or an environmental insult from natural or malicious disasters. Further, current modalities of biotechnology engineering can modify the epigenome at specific sites using engineered molecules such as the  Clustered Regulatory Interspaced Short Palindromic Repeat (CRISPR)-Cas system~\cite{crispr-2017}, which functions as a DNA site-specific nuclease. Here, the very property of fault tolerance of the epigenome may in fact make it harder to engineer persistent changes in the epigenome to effect changes in phenotype or physiological functions. This is desired because without these checks and balances, working in many-to-many capacities and in three-dimensional spatial conformations, the regulation of the genome would be fragile and less efficient or in other words not resilient to small ``bugs'' in the genomic code or environmental assaults. 
Another dimension that results in resilience is the presence of genetic hubs, small class of genes that affect many different biological pathways. 
These hubs tend to have a high level of connectivity in biological networks, and this means that they tend to be a part of large-sized modules and also appear in a large number of modules. On the other hand, the more common non-hub genes form smaller modules and appear in a few modules. 
In network theory, a hub is a node with the number of edges being vastly greater than the average, which is a consequence of the scale-free property of networks~\cite{barabasi2009scale}, as opposed to random networks. 
The spike of hubs in scale-free networks comes from the power-law distribution. Hubs have a significant impact on the network topology. Hubs can also be found in many real networks, most popularly in the internet. It is the presence of these hubs that makes these scale-free networks, which abound everywhere (\eg gene regulatory networks or GRNs, airline traffic routes, weblog links) resilient to random attacks. However, the very presence of hubs can make these networks vulnerable to targeted (hub) attacks. One way to imbue resilience to these targeted attacks could be to adaptively change the networked state from random to differential degrees of scale-freeness based on the intensity and scale of the attack, \eg localized or planet-scale.
\bigskip

\section{Concluding Thoughts}
\label{sec:conclusion}
With the plethora of IoT devices and the rising criticality of the data they generate, it becomes important to ensure the resilience of CPS. This is particularly true for CPS that are being deployed for safety-critical operations and in a subset of these the decisions have to be made on the fly. In addition, for sensitive healthcare data, computation close to the source of the data will help with meeting the privacy guarantees needed for such data. Thus, there are myriad CPS applications with varied resilience requirements. In a lot of these applications, it is the combination of the device-level and the algorithmic resilience that will determine the success of the application. Think of implantable systems for example that are used in the healthcare industry. For those, not only is the device-level security critical but the storage, computation, and transmission of results from the computation, all need to happen fast and in a secure fashion. In contrast, think of the sensors that are embedded in a digital farm in a connected mesh network. There, the latency requirement to get parameters such as moisture and nitrate level are much longer. However, the volumes of noisy data sets and the lack of reliable wireless connectivity pose a distinct set of challenges. 
On an orthogonal track, we discussed the two ways of thinking about resilience---resilience-by-design and resilience-by-reaction. The former encompasses various design and development techniques used to ensure resilience. The latter refers to techniques that are invoked at runtime, during operation of the CPS, to ensure resilience. The two are connected in the sense that today's CPS that is resilient-by-design was designed using ``reactive algorithmic patches'' and were initially resilient-by-reaction. 

In conclusion, we picked a select swath of emerging technologies that can inform the design of new resilient-by-design CPS. These included learning from the resilience of living cells in that the gene regulatory networks within a living cell with its hub-and-spoke structure make the living cells resilient to random attacks, albeit, vulnerable to targeted hub removal through the prior knowledge of the network topography. To circumvent this, we may have to design adaptive networks that oscillate between random and scale-free networks. Another important point is the resilience of data sharing repositories, often operating in a federated manner, with respect to storage and in more advanced federation, computation to fully use the power of the federated servers. Finally, flexible decomposition of computing among sensors, edge, and the cloud is often needed to ensure resilient operation in CPS. An important example of this is object detection from video captured by drones that can happen partly on the drone, partly on nearby edge platforms, and partly communicated back to the cloud platform. Further, the algorithm for the object detection workload may itself be approximated to accelerate detection with comparable accuracy.

\section*{Acknowledgments}

This work is supported in part through NSF Award for the Grand Challenges in Resilience Workshop at Purdue March 19--21, 2019 through award number 1845192. Any opinions, findings, and conclusions or recommendations expressed in this material are those of the authors and do not necessarily reflect the views of the National Science Foundation. 

\bibliography{schaterji,oishi,ResilienceMou,references,sbagchi,suman,game_theory_refs}
\end{document}